\documentclass[conference]{IEEEtran}
\IEEEoverridecommandlockouts
\usepackage{cite}
\usepackage{amsmath,amssymb,amsfonts}
\usepackage{algorithmic}
\usepackage{graphicx}
\usepackage{textcomp}
\usepackage{xcolor}
\usepackage{comment}
\def\BibTeX{{\rm B\kern-.05em{\sc i\kern-.025em b}\kern-.08em
    T\kern-.1667em\lower.7ex\hbox{E}\kern-.125emX}}
\begin{document}

\title{Less Is More? When Dataset Context Hurts LLM-Generated Dataset Descriptions

\thanks{Funded by Siemens AG and the Technical University of Munich – Institute for Advanced Study (TUM-IAS), Germany.}
}

\author{
\IEEEauthorblockN{Anonymous Author(s)}
\IEEEauthorblockA{Affiliation(s) withheld for double-blind review}
}
\author{\IEEEauthorblockN{1\textsuperscript{st} Lisa-Yao Gan}
\IEEEauthorblockA{
\textit{Technical University of Munich}\\
Munich, Germany\\
lisa.gan@tum.de}
\and
\IEEEauthorblockN{2\textsuperscript{nd} Arunav Das}
\IEEEauthorblockA{
\textit{King's College London}\\
London, United Kingdom \\
arunav.das@kcl.ac.uk}
\and
\IEEEauthorblockN{3\textsuperscript{rd} Johanna Walker}
\IEEEauthorblockA{
\textit{King's College London}\\
London, United Kingdom \\
johanna.walker@kcl.ac.uk}
\and
\IEEEauthorblockN{4\textsuperscript{th} Klaus Diepold}
\IEEEauthorblockA{
\textit{Technical University of Munich}\\
Munich, Germany\\
kldi@tum.de}
\and
\IEEEauthorblockN{5\textsuperscript{th} Elena Simperl}
\IEEEauthorblockA{
\textit{King's College London}\\
London, United Kingdom \\
\textit{TUM Institute for Advanced Study (TUM-IAS)} \\
\textit{Technical University of Munich}\\
Munich, Germany \\
elena.simperl@kcl.ac.uk}
}
\maketitle

\begin{abstract}
Dataset search and reuse are strongly constrained
by the quality of metadata such as natural language descriptions,
which are often sparse or inconsistent. Although large language
models (LLMs) can generate such descriptions automatically,
little empirical guidance exists on what makes a good dataset
description and what dataset context LLMs actually need. We
study these questions through a literature-grounded framework
of dataset description quality and a large-scale ablation study
using 252 datasets (1,336 CSV files) from the European data
portal data.europa.eu. We generate descriptions with LLMs
in a baseline scenario and two ablation scenarios: (1) using
only dataset titles, (2) titles and schema, and (3) titles, schema
and representative data, and evaluate them with an LLM-as-a-
judge framework and a semantic descriptive attribute analysis
grounded in our quality dimensions. Our results reveal a consis-
tent schema penalty: table-schemas alone often degrade narrative
quality, while representative data partially restores grounding
without improving overall human-facing quality. We further
show that different LLMs exhibit stable descriptive personas.
These findings provide practical guidance for LLM-supported
data publishing workflows.
\end{abstract}


\section{Introduction}
The ability to identify and locate appropriate data is fundamental to its reuse. Portals are a key way in which data publishers have facilitated this. Platforms such as data.europa.eu, national open government portals, and institutional repositories host millions of datasets \cite{EC_OpenDataGuide_2024}. And yet users frequently report difficulties both in locating data that match their information needs and in understanding the datasets they encounter. Prior work consistently shows that these difficulties are fundamentally constrained by the quality of dataset metadata , which are often sparse, inconsistent, or poorly written \cite{Chapman2020-sy, Gregory2020-mg, 10.1145/3665939.3665959}. Common metadata for datasets include the title, tags and descriptions. As portal search is primarily keyword based, poorly-written descriptions represent a missed opportunity to match the needs of users. At the same time, empirical studies of dataset discovery and reuse consistently show that natural-language overview descriptions are among the most important metadata elements supporting user sensemaking, relevance assessment, and reuse decisions \cite{Mathiak-2023, KOESTEN2021102562, Gregory2020-mg, 10.1371/journal.pone.0246099, info:doi/10.2196/25440}. As a result, users struggle not only to retrieve relevant datasets, but also to interpret their contents, assess fitness for use, and build trust in unfamiliar data \cite{10.1145/3025453.3025838, KOESTEN2021102562}.

Together, this literature establishes (the lack of) dataset description quality as a central bottleneck in data discovery and documentation.

Motivated by these persistent problems, recent research has begun to explore the use of large language models (LLMs) to automate dataset documentation and description generation \cite{zhang2025autoddgautomateddatasetdescription}. These efforts demonstrate that LLMs can generate coherent natural-language summaries and substantially expand metadata coverage. However, from the perspective of data publishers and portal operators, a critical practical question remains unanswered: \emph{what information is actually necessary to provide to an LLM in order to reliably generate high-quality dataset descriptions?}
Dataset providers often lack the time and incentives to curate rich metadata, motivating workflows where a dataset title alone could yield a useful description. While additional signals such as schema or data samples may help, they also introduce noise: too little context risks vague summaries, while too much may overwhelm the model. There is little empirical guidance on where this “sweet spot” lies between insufficient and excessive dataset context.

In this work, we address this gap through two complementary research questions:
\begin{itemize}
    \item \textbf{RQ1:} To what extent do LLM-generated dataset descriptions meet the characteristics of high-quality dataset descriptions?
    \item \textbf{RQ2:} How does the quality of LLM-generated dataset descriptions vary under different dataset-context prompting conditions?
\end{itemize}

To answer \textbf{RQ1}, we synthesize prior research on dataset discovery, metadata quality, and data sensemaking to derive a structured characterization of high-quality dataset descriptions. To answer \textbf{RQ2}, we conduct a large-scale ablation study, examining how description quality changes as progressively richer dataset context is provided. We evaluate generated descriptions using both quality scoring and semantic descriptive analysis.

Our work makes three primary contributions:
\begin{itemize}
    \item \textbf{Literature-grounded characterization.} We consolidate prior research into a structured framework of what constitutes a high-quality dataset description.
    \item \textbf{Ablation study of LLM-based description generation.} We provide an empirical analysis of how description quality changes as increasingly rich dataset signals are provided to an LLM.
    \item \textbf{Practical guidance for data publishers.} We derive empirically grounded insights into what dataset information is most valuable to provide when using LLMs to automatically generate dataset descriptions.
\end{itemize}

This work aims to clarify both what makes dataset descriptions effective for users and how data publishing workflows can best leverage LLMs to generate such descriptions automatically at scale.

\section{Related Work}

Our work builds on prior research in dataset discovery and on recent efforts to apply LLMs to dataset documentation and metadata enrichment. We review these two lines of work and position our contribution relative to them.

\subsection{Dataset Discovery and the Role of Descriptive Metadata}

A substantial body of research characterizes dataset discovery as exploratory, iterative, and fundamentally different from traditional document retrieval \cite{Chapman2020-sy, Gregory2020-mg, Walker2024DataPrompting}. These studies show that dataset search is shaped by ambiguous information needs and heavy reliance on contextual cues, making descriptive metadata central to relevance assessment and sensemaking.

Within this literature, natural-language descriptions consistently emerge as a primary mechanism for interpreting unfamiliar datasets, assessing trustworthiness, and evaluating fitness for use \cite{KOESTEN2021102562, Mathiak-2023, 10.1371/journal.pone.0246099}. Recent work further shows that narrative description fields dominate natural-language and conversational dataset retrieval, and that keyword-style metadata alone is insufficient to capture users’ dataset needs \cite{gan2025keywordskeymetadatafield}. Their results also indicate that enriched, LLM-generated descriptions can substantially improve retrieval effectiveness.

\subsection{LLMs for Dataset Description and Metadata Enrichment}

Generative models are increasingly explored as practical tools for automating and enriching dataset metadata. Recent work shows that LLMs support a wide range of metadata tasks, including automated annotation, enrichment, normalization, and retrieval-oriented description generation \cite{Yang2025-kd, zhang2025autoddgautomateddatasetdescription, 10.1093/bioinformatics/btaf519, Ahmed_Polini_2025, Turner2025-qf, 10.1145/3705328.3748100}. Collectively, these studies demonstrate that structured pipelines, retrieval augmentation, and model adaptation can improve the reliability and downstream utility of LLM-generated metadata.

\section{Key Characteristics of a Good Dataset Description}
Drawing from recent literature and open data guidelines, we identify several key characteristics that constitute a high-quality dataset description. These characteristics are not only important for enhancing dataset discoverability, especially in natural language search settings, but also crucial for supporting users in understanding and making use and sense of the data.
Table~\ref{tab:desc_chars} summarises the characteristics of high-quality dataset descriptions.

\begin{table*}[!htb]
\caption{Literature-grounded characteristics of high-quality dataset descriptions}
\begin{center}
\begin{tabular}{|p{3.2cm}|p{7.2cm}|p{5.0cm}|}
\hline
\textbf{Characteristic} & \textbf{What the description should convey} & \textbf{Example cues (signals)} \\
\hline
Overview \& purpose &
Plain-language summary of what the dataset is about and why it exists (intended analytical or policy purpose). &
Topic/domain, phenomenon, intended use or goal \\
\hline
Contents \& coverage &
What variables/fields the dataset contains and what they represent, including spatial and temporal scope and relevant granularity. &
Key attributes, units, geography, time range, resolution \\
\hline
Structure \& size &
How the dataset is organized and delivered, including format and basic scale/complexity. &
File/API type, rows/records, columns/attributes, multiple files, nested structures \\
\hline
Provenance \& updates &
Who produced the data and how current it is, including publication date and update frequency. &
Publisher/source, collection process, last updated, update schedule \\
\hline
Quality \& limitations &
Known issues, uncertainty, and methodological caveats that affect interpretation and reuse. &
Missingness, bias, measurement changes, known errors, comparability notes \\
\hline
Usage notes \& insights &
How the dataset can be used and what it may reveal, including suggested use cases or notable patterns. &
Example analyses, intended applications, notable trends/anomalies \\
\hline
Clarity \& plain language &
Accessible writing that avoids unexplained jargon, acronyms, or insider terminology. &
Definitions, expanded acronyms, simple phrasing, self-contained explanation \\
\hline
User vocabulary alignment &
Terms that match how users search, including synonyms and related phrases to reduce vocabulary mismatch. &
Common query terms, synonyms, abbreviations, alternative names \\
\hline
\end{tabular}
\label{tab:desc_chars}
\end{center}
\end{table*}

\subsection{Clear Overview and Purpose}
First and foremost, an effective dataset description should provide a \textbf{clear overview and purpose}. This includes a concise summary of what the dataset is about and why it exists, ideally expressed in plain language. For example, a good description might read: \textit{“This dataset contains annual city-level crime statistics in London, collected to analyse trends in street crime over the past decade.”} Recent empirical analysis of researcher needs confirms that a "better overview" remains a top user requirement for framing data discovery \cite{Mathiak-2023}. Beyond basic discovery, Pentz et al. \cite{Pentz2025Metadata2050} argue that these descriptive summaries serve as critical "trust signals" within a modern "Research Nexus," providing the nuanced context necessary to assess a dataset's credibility and "contextual fitness" for reuse. Furthermore, the use of plain language—specifically prioritizing simplification and informativeness—is increasingly recognized as a core metric for making such scientific summaries accessible and trustworthy for non-expert audiences \cite{guo-etal-2024-appls}.

\subsection{Contents and Coverage}
In addition to this high-level overview, the description should include details about the \textbf{contents and coverage} of the dataset. This involves naming the main variables or fields, what they represent, and the spatial or temporal scope of the data. Users benefit from knowing, for instance, whether the dataset covers all UK cities from 2010 to 2020, or just selected regions in a given year. Including this information helps users quickly assess relevance based on their specific data needs \cite{10.1145/3746059.3747727}. Empirical analysis of user data requests confirms that geospatial and temporal attributes, particularly the required level of granularity, are the most critical features users seek when evaluating a dataset’s contents and coverage \cite{KACPRZAK201937, Chapman2020-sy}.

\subsection{Structure and Size}
Another essential characteristic is the \textbf{structure and size} of the dataset. A good description should communicate key structural properties such as the format (e.g., CSV, JSON, API), number of records, and number of attributes or columns. Surfacing this information in the dataset’s natural-language description allows users to rapidly assess technical suitability and anticipated effort, and reflects widely recognized requirements for effective dataset documentation and reproducible computational research \cite{LEIPZIG2021100322, Canham2016-hs, info14080427}. Evidence from large-scale log analyses of national data portals further shows that users actively seek this type of structural information during dataset search: file format is a primary filter, and a substantial proportion of external queries explicitly include technical extensions such as “CSV” or “JSON” to ensure immediate usability \cite{KACPRZAK201937}.
For example, indicating that a dataset includes 10 columns and approximately 5,000 rows provides an immediate sense of its granularity and potential usability, while noting whether it is distributed across multiple files or contains nested structures further clarifies the technical complexity involved. Such structural cues support early-stage sensemaking and help users anticipate the practical demands of working with the data \cite{KOESTEN2020102367, https://doi.org/10.1002/asi.24961}.

\subsection{Provenance and Update Information}
Furthermore, dataset descriptions should specify \textbf{provenance and update information}. This includes identifying the source organization, the date of publication, and the update frequency, because users need to know where the data originated and how current it is to build trust and assess its suitability for reuse. Provenance metadata documents the processes that produced the data and is a recognized component in metadata standards that support discovery, reuse, and reproducibility \cite{LEIPZIG2021100322, info14080427, 10.1098/rsta.2021.0300}.
Surfacing this information directly in the dataset’s natural-language description, for example, through simple factual statements such as \textit{“Data provided by the London Metropolitan Police; last updated June 2025 (updated annually)”}, helps users interpret the dataset’s lineage and currency. These are critical cues in judging data quality and reliability \cite{FanielYakel2017DataReuse, doi:10.1177/0165551519837182, KOESTEN2021102562}.

\subsection{Quality and Limitations}
To further support user interpretation, descriptions should openly discuss \textbf{quality and limitations}. This includes information about missing data, inconsistencies, methodological notes, or caveats in the data collection process. For example, if some crime locations were not recorded—resulting in 5\% missing values—that should be stated explicitly in the description. Likewise, if data collection methods changed mid-series, this should be flagged, as it may affect longitudinal comparability.

Koesten et al. demonstrate that data sensemaking depends on understanding how data were produced, what uncertainties or problems they may contain, and what limitations shape their interpretation \cite{KOESTEN2021102562, https://doi.org/10.1002/asi.23730}. Large-scale survey evidence further shows that researchers consider information about quality, trust, and data issues to be central when evaluating datasets and deciding whether they are suitable for reuse \cite{Gregory2020-mg}.
Research on data reuse reinforces this need for transparency: Pasquetto et al. show that successful reuse requires access to documentation about data production, processing, and limitations, and that the absence of such information makes datasets difficult to interpret and trust. Borgman similarly argues that data removed from their original context are not self-explanatory, and that understanding assumptions, uncertainties, and quality constraints is essential for responsible interpretation and reuse \cite{Pasquetto-2017, https://doi.org/10.1002/asi.22634}.

\subsection{Usage Notes and Potential Insights}
Where possible, dataset descriptions should also include \textbf{usage notes and potential insights}. These can take the form of suggested use cases or brief summaries of preliminary findings derived from the dataset. For example, a description might note that the data could be used to evaluate the impact of policy interventions on crime rates or to perform spatial analyses across boroughs. Even brief hints about notable trends or anomalies—such as a spike in incidents during a specific year—can spark ideas for further analysis and support early-stage exploration.

Research on exploratory information seeking emphasizes that users often approach complex resources not only to retrieve known facts, they also build understanding, generate hypotheses, and identify promising analytical directions. Therefore, summaries that highlight potential interpretations and avenues for inquiry are critical in supporting this process \cite{10.1145/1121949.1121979}. For instance, users also rely on rich descriptive context to infer what kinds of questions a dataset might help answer \cite{kups9359}. In parallel, research on data reuse demonstrates that once data are detached from their original production contexts, interpretive cues and examples become important mechanisms for making data intelligible and actionable for new users \cite{10.1111/j.1083-6101.2007.00342.x}.

\subsection{Clarity and Plain Language}
In terms of writing style, the dataset description should prioritize \textbf{clarity and plain language}. Descriptions should avoid unexplained acronyms, technical jargon, or internal project terminology that outsiders may not understand. For example, instead of writing “hydro infrastructure geospatial data,” it is clearer to say “GPS coordinates of public water fountains.” Writing in an accessible, self-contained manner lowers the barrier to entry for non-experts and supports early-stage sensemaking.

Large-scale analysis of data repositories shows that the interpretability and readability of the dataset description text itself significantly influence user engagement: clearer, more readable descriptions are associated with higher dataset downloads, while overly complex or dense descriptions deter use \cite{Liu2024-zu}. Qualitative studies of data discovery and sensemaking further demonstrate that users strongly depend on natural-language descriptions to understand what a dataset contains, and frequently experience frustration, uncertainty, and additional effort when descriptions are unclear, overly technical, or insufficiently explained \cite{KOESTEN2021102562, 10.1145/3025453.3025838}. 

\subsection{Alignment with User vocabulary}
Finally, descriptions should exhibit strong \textbf{alignment with user vocabulary}. This means using words and phrases that people are likely to use in search queries \cite{soton375700}. A good practice is to incorporate synonyms or related phrases. For instance, a dataset about “automobile accidents” might also mention “road incidents” or “traffic collisions” to ensure it is findable by a wider range of search terms.

Research on open data discovery highlights that when descriptive metadata relies on narrow or inconsistent terminology, relevant datasets are often missed due to vocabulary mismatches between data publishers and users. Křemen and Nečaský show that weak or shallow descriptions and inconsistent vocabularies directly undermine discoverability, and that aligning descriptive metadata with the terms users employ is necessary to reduce false negatives in dataset search \cite{KREMEN20191}. Log analyses of dataset search behaviour further demonstrate that users frequently search using abbreviations, acronyms, and everyday terms, and that failing to reflect this language in descriptions leads to missed retrieval opportunities \cite{KACPRZAK201937}. 

In summary, a good dataset description acts as both a discovery tool and a sensemaking aid. It bridges the gap between raw data and user needs by combining content-specific, structural, contextual, and linguistic features in a concise yet rich narrative.

\section{Methodology}
\begin{figure*}[!htb]
    \centering
    \includegraphics[width=0.75\textwidth]{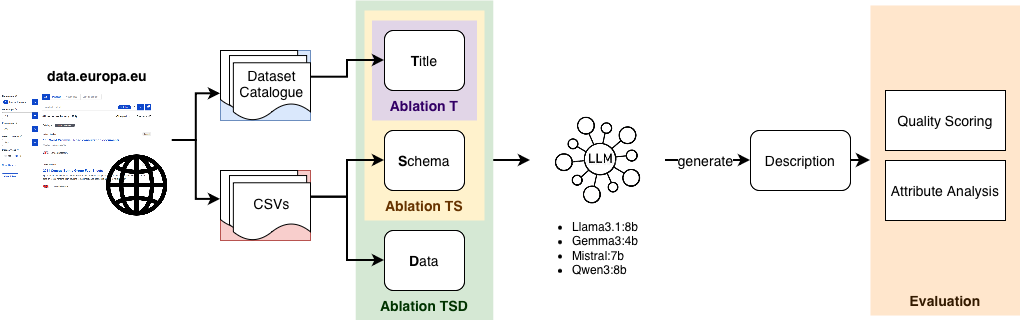}
    \caption{Overview of the experimental methodology and ablation design. We construct progressively richer dataset representations from open data catalogues and CSV resources (title, schema, and samples), use these as input to an LLM to generate dataset descriptions, and evaluate outputs using quality scoring and descriptive attribute analysis.}
    \label{fig:methodology_overview}
\end{figure*}

We investigate how the quality of LLM-generated dataset descriptions varies with the amount of dataset context provided. We design an ablation study grounded in a realistic open data publishing workflow, constructing progressively richer dataset representations (Fig.~\ref{fig:methodology_overview}) and evaluating the resulting descriptions using quality scoring and descriptive attribute analysis. By “richer” dataset representations, we refer to incrementally adding more dataset context. In detail: moving from titles only to including schema and representative data.

\subsection{Dataset Collection}
We conduct our experiments on datasets from the London Datastore (LDS), a major open government data portal indexed by data.europa.eu. LDS provides a realistic open-data testbed in English with broadly comparable metadata practices, while still exhibiting the incompleteness and variability typical of large public data portals.


Using the public catalogue API, we retrieved the full catalogue and filtered to datasets containing at least one downloadable CSV resource, yielding 252 datasets comprising 1,336 CSV files. Many datasets include multiple CSV resources corresponding to different years, geographic partitions, or related subtables. We preserve the dataset-level grouping defined by the portal and exclude datasets without readable tabular content after parsing.

\subsection{Dataset Preprocessing and Representation}

For each dataset, we construct a bounded, model-agnostic representation capturing three signals: title, structural schema, and representative data samples. All CSV files are parsed using a robust reader, and up to the three largest tables per dataset are retained. For each table, we extract column headers (capped at 20) and representative samples (first three and last three rows). Cell values are lightly normalized by removing line breaks and truncating long strings. This yields a unified dataset snapshot comprising a title and up to three tables, each represented by its schema and example records.

\subsection{Baseline and Ablation Design: Dataset Context Conditions}
We define a title-only baseline and two progressively richer conditions:

\begin{enumerate}
    \item \textbf{Title-only (T):} dataset title only.
    \item \textbf{Title + Schema (TS):} title and column headers.
    \item \textbf{Title + Schema + Data (TSD):} title, column headers, and representative data rows.
\end{enumerate}

Across all conditions, the system persona and task instruction remain fixed. No templates, quality criteria, or few-shot examples are used. For multi-table datasets, schema and samples are presented in a structured table-by-table format with fixed limits to control prompt size.

\subsection{Description Generation and Prompting Setup}

We generate descriptions using several open-weight instruction-tuned models executed locally via Ollama (LLaMA-3.1-8B, Qwen-3-8B, Mistral-7B, Gemma-3-4B). These models were selected to reflect a diverse set of widely used, openly available LLMs with relatively low computational requirements. Each dataset is processed sequentially, and all ablations are generated from the same dataset snapshot.

We use a fixed prompt structure consisting of a system role defining the model as an expert dataset cataloguer, a task instruction to produce an open-data-style dataset description, and a dataset context block containing the condition-dependent inputs (Fig.~\ref{fig:prompt}).

\begin{figure}[!htb]
\centering
\fbox{
\begin{minipage}{0.95\columnwidth}
\footnotesize
\textbf{System:} You are an expert database cataloguer, data engineer, and knowledge scientist specializing in urban and rural datasets, particularly for London and the UK. Your role is to create dataset descriptions for an open data catalogue based on the information provided.

\vspace{0.5em}
\textbf{User:} When given dataset-related content such as a dataset title, column names, or example records, produce a dataset description suitable for an open data catalogue. 

\vspace{0.5em}
\textbf{Dataset context:}  
[Dataset title]  
[Condition-dependent: table schemas]
[Condition-dependent: representative data rows]

\end{minipage}
}
\caption{Prompt structure used across all experiments. The only variation across ablation conditions is the dataset context provided.}
\label{fig:prompt}
\end{figure}

\subsection{Evaluation Overview}
We evaluate generated descriptions using two complementary approaches: (1) LLM-as-a-judge quality scoring grounded in our literature-derived characteristics, and (2) descriptive attribute analysis examining how models redistribute descriptive focus as context increases. We additionally compare generated descriptions against original publisher-provided descriptions using the same quality rubric.

\subsubsection{LLM-as-a-Judge Quality Evaluation}

We employ an LLM-as-a-judge framework to systematically score the quality of generated dataset descriptions under each ablation condition. An open-weight judge model (gpt-oss:20b) is prompted to act as an expert dataset cataloguer and evaluate each description using a detailed rubric aligned with our eight literature-grounded characteristics: Overview \& Purpose, Contents \& Coverage, Structure \& Size, Provenance \& Updates, Quality \& Limitations, Usage Notes \& Insights, Clarity \& Plain Language, and User Vocabulary Alignment.

Each characteristic is scored on a 1--5 Likert scale, and the judge is required to provide a brief justification alongside each numeric score. All evaluations are returned in structured JSON format and aggregated at the dataset and condition level.

To isolate the effect of additional dataset context, we compute three derived metrics: (1) \textit{Schema effect}, measuring the score change from Title-only to Title+Schema; (2) \textit{Data effect}, measuring the marginal change from Title+Schema to Title+Schema+Data; and (3) \textit{Net effect}, measuring the total change from Title-only to Title+Schema+Data. 

In addition, to contextualize LLM-generated descriptions relative to existing human-authored metadata, we conduct a secondary evaluation comparing descriptions generated under the full-context condition (Title+Schema+Data) against the original publisher-provided catalogue descriptions. Using the same LLM-as-a-judge rubric and scoring procedure, we assess whether LLM-generated descriptions score higher, equal, or lower than the human-authored descriptions for each dataset.

We note that this evaluation uses an LLM-as-a-judge rather than human annotators. While this enables large-scale and consistent scoring, it cannot substitute for human-centered evaluation and may reflect biases of the judge model. Therefore, our results should be interpreted as comparative signals across conditions rather than absolute measures of description quality. More specifically, these comparisons reflect the preferences of the judge model and do not necessarily transfer to human evaluators. In particular, when comparing LLM-generated descriptions to publisher-/human-authored catalogue metadata, the judge may systematically prefer LLM-style writing—even in cases where a human reviewer would rate the human-authored metadata as better. We will conduct user studies in the future.

\subsubsection{Descriptive Attribute and Structural Focus Analysis}
To complement scalar quality scores, we conduct a descriptive attribute analysis that quantifies how models redistribute emphasis across our eight characteristics as context increases. For each generated description, we segment the text into \emph{descriptive units}, defined as sentence- or bullet-level propositions obtained by splitting on punctuation and list markers. We then apply lightweight, hand-written regular-expression rules to retain units that express explicit descriptive content (e.g., schema cues such as ``column'', ``field'', ``type'', ``table''; provenance cues such as ``source'', ``published'', ``updated''; and quality/usage cues such as ``missing'', ``limitations'', ``used for''). Each retained unit is \emph{semantically normalised} by lowercasing and lemmatising simple surface forms (e.g., plural handling), and is mapped to exactly one of the eight characteristic categories using category-specific keyword sets and pattern matches, with deterministic tie-breaking rules.
We report the proportional distribution of category assignments per model and ablation condition to characterise how additional context shifts descriptive focus. In addition, we compute two derived measures: (i) \emph{knowledge intensity}, operationalised as the number of retained descriptive units per description (optionally normalised by description length), and (ii) \emph{structural volatility}, operationalised as the number of distinct sub-attributes surfaced per condition, where a sub-attribute corresponds to a de-duplicated unit ``header'' (i.e., a normalised attribute phrase or schema/value facet) extracted from the retained units. This analysis reveals whether added context primarily increases narrative clarity, shifts descriptions toward technical structure, foregrounds administrative provenance, or surfaces more fine-grained analytical statements.

\section{Results}
We present results from two complementary analyses. First, we report LLM-as-a-judge quality evaluations to quantify how overall description quality changes across ablation conditions. Second, we analyse structural shifts in descriptive focus to examine how models redistribute attention across different description characteristics as dataset context increases.

\subsection{LLM-as-a-Judge Quality Outcomes}

The LLM-as-a-judge evaluation reveals a consistent behavioural pattern across all tested models (LLaMA-3, Qwen, Mistral, Gemma). Introducing technical schema information without data samples leads to a negative marginal effect on overall description quality for every model. This schema penalty is strongest for Qwen ($-0.221$) and Mistral ($-0.158$), followed by Gemma ($-0.136$), and is weakest for LLaMA-3 ($-0.073$). This degradation primarily affects narrative-oriented characteristics, particularly Overview \& Purpose and Clarity \& Plain Language, indicating that exposure to column headers alone systematically shifts model outputs toward technical enumeration at the expense of user-centered description.

Adding representative data samples produces a positive marginal effect across all models, partially offsetting the schema penalty. However, the magnitude of this data recovery effect is consistently smaller than the preceding decline. LLaMA-3 shows only a modest improvement ($+0.031$), while Qwen exhibits almost no recovery ($+0.022$). Mistral ($+0.060$) and Gemma ($+0.115$) benefit more strongly from the introduction of data samples, with Gemma showing the largest marginal gain. Despite these increases, in most cases the full context condition (Title+Schema+Data) does not surpass the Title-only baseline in overall quality, indicating that additional technical grounding does not automatically translate into better human-facing descriptions. This counterintuitive pattern likely reflects a \emph{conditioning effect} rather than a general advantage of less context: exposing models to schema (and, to a lesser extent, sample rows) can shift generation toward low-level structural enumeration and away from higher-level synthesis (e.g., overview, audience framing, caveats) that our rubric rewards. Conversely, Title-only prompts may elicit fluent, complete-sounding narratives that can include plausible but \emph{unsupported} details (e.g., update cadence or methodological caveats) that are not recoverable from the title alone. As a result, our LLM-as-a-judge scores should be interpreted as \emph{comparative signals of perceived description quality under the judge model}, not as verified factual correctness when the necessary evidence is absent from the input.

\begin{figure}[htbp]
\centering
\includegraphics[width=\columnwidth]{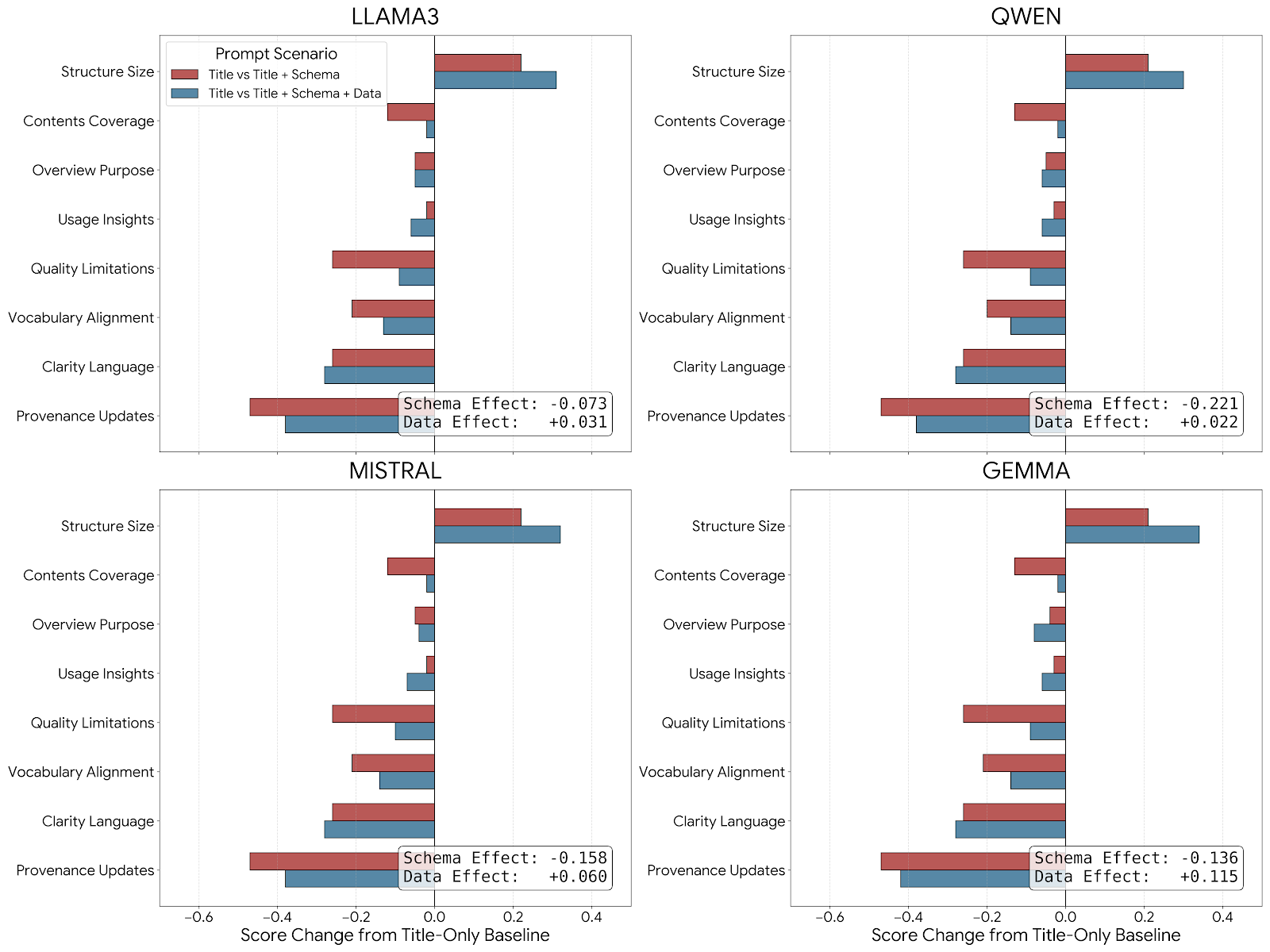}
\caption{Characteristic-level quality changes relative to the Title-only baseline under different prompt conditions, evaluated using an LLM-as-a-judge framework. Red bars show the effect of adding schema (Title+Schema), while blue bars show the marginal effect of adding representative data samples (Title+Schema+Data). Across all models, adding schema introduces a consistent penalty, particularly for Overview \& Purpose, Clarity \& Plain Language, and Provenance \& Updates. Adding data partially recovers quality, with the strongest gains observed for Structure \& Size.}
\label{fig:judge-characteristics}
\end{figure}

Table~\ref{tab:judge-effects} summarises the average overall quality changes across models, highlighting both the consistent schema penalty ($-0.073$ to $-0.221$) and the smaller but positive marginal contribution of data samples ($+0.022$ to $+0.115$). Fig.~\ref{fig:judge-characteristics} further shows that the schema penalty is strongest for narrative-oriented characteristics, while representative data samples most strongly benefit Structure \& Size and Contents \& Coverage.

At the characteristic level, Structure \& Size exhibits the strongest net improvement across conditions, while Provenance \& Updates and Clarity \& Plain Language remain consistently vulnerable to the introduction of schema. Together, these results demonstrate that richer dataset context acts as a double-edged sword: it improves technical descriptiveness and factual grounding, but often weakens narrative accessibility and administrative framing unless these aspects are explicitly reinforced.

\begin{table}[htbp]
\caption{Average overall quality score changes relative to the Title-only baseline. ‘Schema Effect’ measures the change from T → TS. ‘Data Effect’ measures the marginal change from TS → TSD.}
\label{tab:judge-effects}
\begin{center}
\begin{tabular}{|l|c|c|}
\hline
\textbf{Model} & \textbf{Schema Effect} & \textbf{Data Effect} \\
\hline
LLaMA-3 & -0.073 & +0.031 \\
\hline
Qwen & -0.221 & +0.022 \\
\hline
Mistral & -0.158 & +0.060 \\
\hline
Gemma & -0.136 & +0.115 \\
\hline
\end{tabular}
\end{center}
\end{table}

To contextualize these effects relative to current practice, we further compared LLM-generated descriptions produced under the full-context condition (TSD), which is our strongest technical grounding setting, to the original publisher-written catalogue descriptions. Across all tested models, generated descriptions more frequently outperform than underperform existing metadata, with ties occurring less often. This indicates that even when schema degrades quality relative to the Title-only baseline, LLM-generated descriptions typically still exceed the quality of current publisher-authored descriptions found in open data portals. Table~\ref{tab:publisher-comparison} summarises these outcomes under the full context condition. However, this comparison should be treated with caution: an LLM judge may systematically prefer longer, more ``complete-sounding'' prose typical of LLM outputs, and may penalize concise publisher metadata even when it is accurate and appropriate for catalogue use. Accordingly, the results in Table~\ref{tab:publisher-comparison} reflect the \emph{preferences of the judge model} rather than definitive evidence that LLM-generated descriptions are superior to human-authored metadata for human readers.

\begin{table}[!htbp]
\caption{Comparison between LLM-generated and publisher-provided (human-authored) dataset descriptions under the full context condition (TSD), illustrating how automatically generated descriptions compare to existing catalogue metadata written by data publishers. Values show the number (and percentage) of datasets for which LLM-generated descriptions score higher, equal, or lower than publisher descriptions under the LLM-as-a-judge evaluation.}
\label{tab:publisher-comparison}
\centering
\renewcommand{\arraystretch}{1.15}
\begin{tabular}{|l|c|c|c|c|}
\hline
\textbf{Model} & \textbf{Higher} & \textbf{Equal} & \textbf{Lower} & \textbf{Total} \\
\hline
Qwen    & 224 (93.3\%) & 6 (2.5\%)  & 10 (4.2\%)  & 240 \\
\hline
LLaMA-3 & 203 (84.6\%) & 4 (1.7\%)  & 33 (13.8\%) & 240 \\
\hline
Gemma  & 154 (77.0\%) & 10 (5.0\%) & 36 (18.0\%) & 200 \\
\hline
Mistral & 132 (73.3\%) & 12 (6.7\%) & 36 (20.0\%) & 180 \\
\hline
\end{tabular}
\end{table}

\subsection{Structural Shifts in Descriptive Focus}
The descriptive attribute analysis reveals that models restructure their descriptive focus in different ways as dataset context increases.

Table~\ref{tab:distribution} summarises the percentage distribution of descriptive sub-attribute categories across models and scenarios, providing an aggregate view of how knowledge emphasis shifts as context increases.

LLaMA-3 exhibits the most stable and human-centric behaviour: Across scenarios, it maintains a consistent emphasis on Overview \& Purpose while increasing its focus on Clarity \& Plain Language when representative data is introduced. 

Gemma demonstrates strong systemic rigidity: The proportional distribution of its descriptive categories remains nearly unchanged across all three scenarios. This reflects a governance-adherent persona that foregrounds administrative lineage and quality cues regardless of the underlying data.

Mistral shows a pronounced reallocation toward structural and relational knowledge: As context increases, its emphasis on Contents \& Coverage rises sharply, indicating a shift toward entity-centric and schema-driven description, which is accompanied by a relative decline in clarity-oriented descriptors.

Qwen exhibits the highest structural volatility. It significantly reduces its focus on Overview \& Purpose as context increases. It starts reallocating its proportional emphasis toward analytical clarity and granular data-driven statements.

\begin{figure}[htbp]
\centering
\includegraphics[width=\columnwidth]{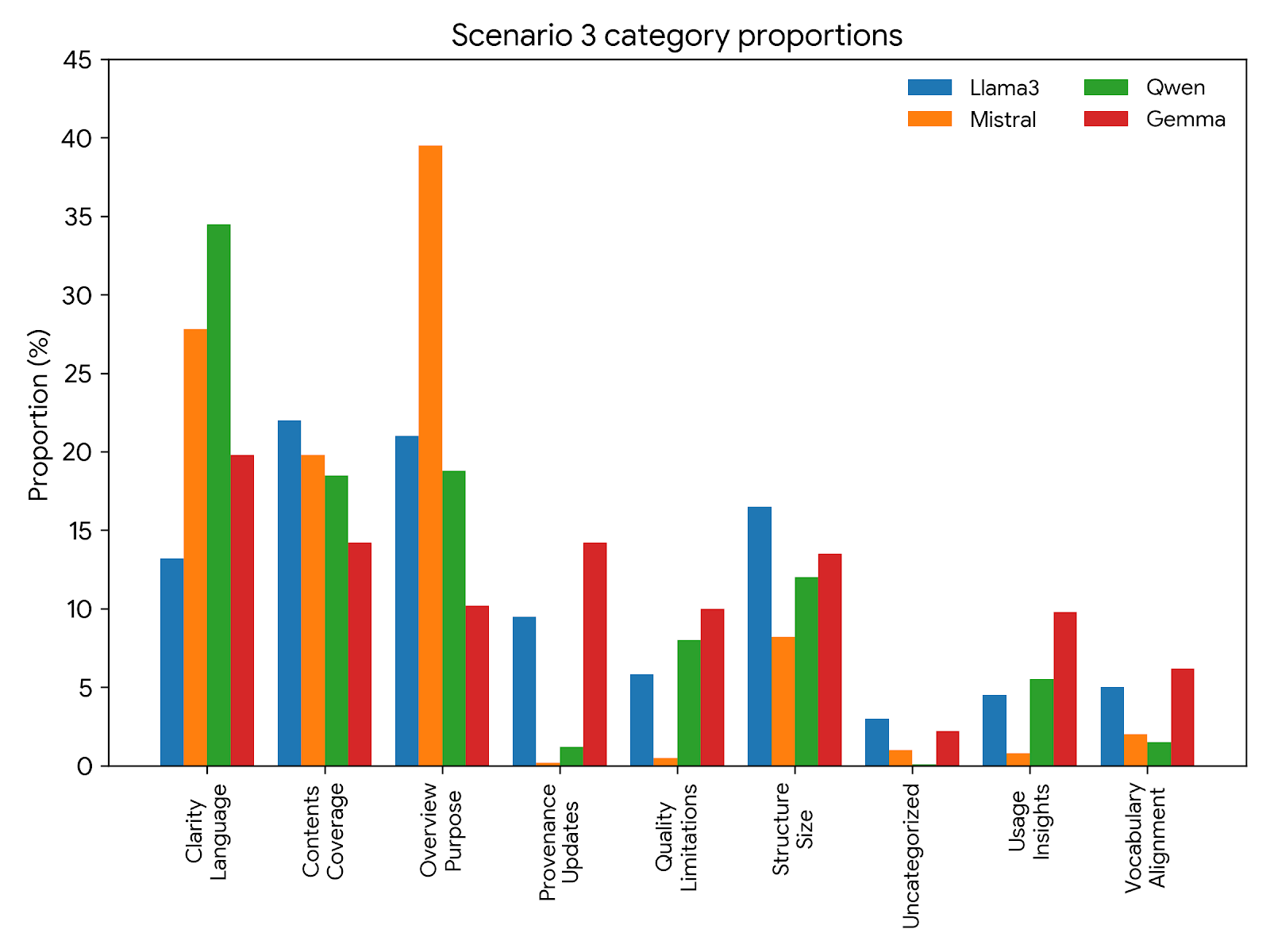}
\caption{Model behavioural archetypes under the full context condition (Title+Schema+Data). The plot shows the proportional distribution of descriptive categories, highlighting distinct descriptive personas across models.}
\label{fig:radar}
\end{figure}

\begin{figure}[htbp]
\centering
\includegraphics[width=\columnwidth]{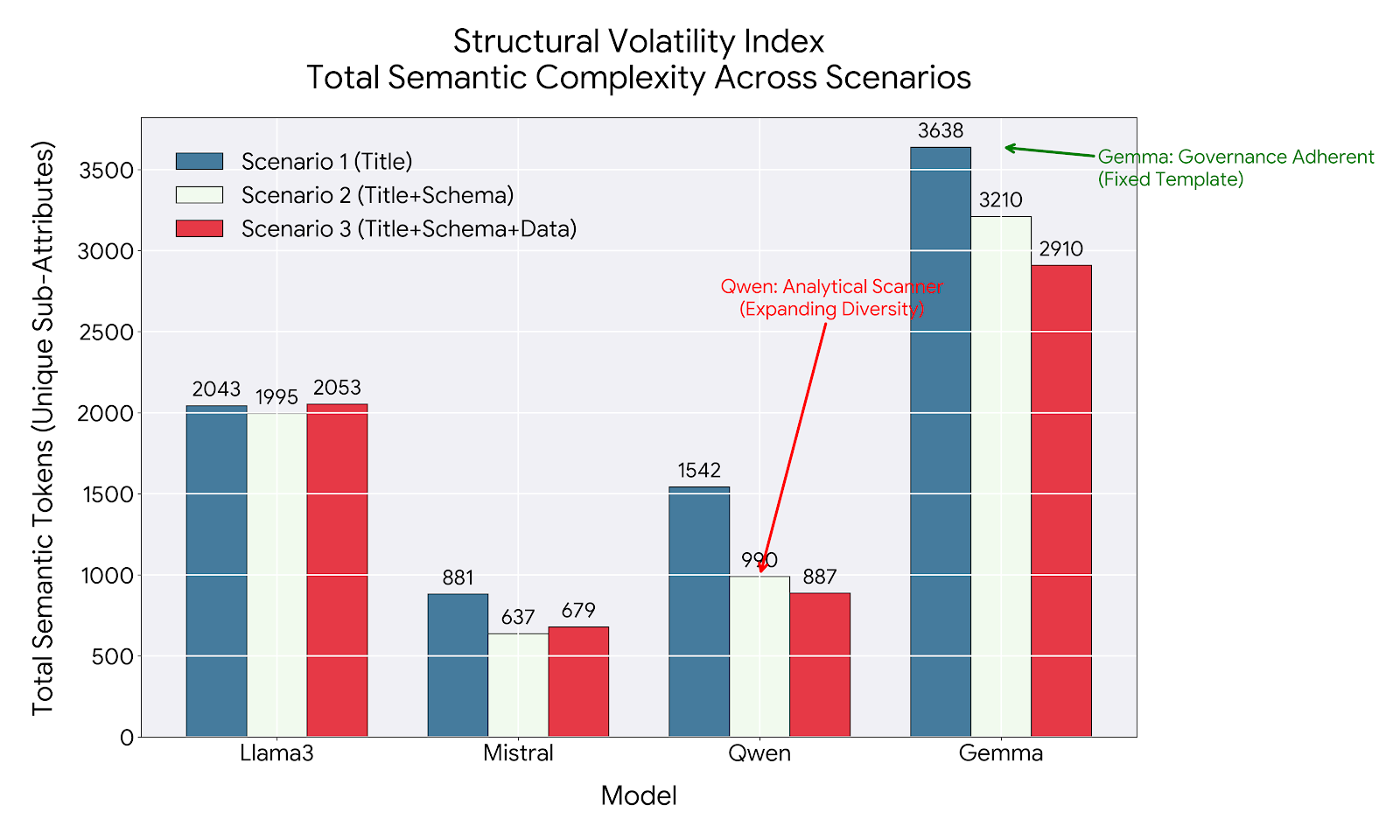}
\caption{Number of unique descriptive sub-attributes discovered per model across scenarios, used as a proxy for structural volatility and sensitivity to new information.}
\label{fig:volatility}
\end{figure}

Fig.~\ref{fig:radar} visualizes the final knowledge representation profiles for each model under the full context condition, highlighting distinct descriptive personas. In contrast, Fig.~\ref{fig:volatility} quantifies how strongly each model restructures the \emph{breadth} of its descriptive attributes across scenarios, measured by the number of unique descriptive sub-attributes surfaced. The plots show that structural volatility generally \emph{decreases} as more context is introduced: Qwen and Gemma exhibit the largest reductions in unique sub-attributes from Title-only to Title+Schema+Data, whereas LLaMA-3 remains comparatively stable across conditions.

Together, these results reveal that increasing context does not simply produce richer descriptions, but gives rise to distinct, model-specific descriptive profiles that determine which forms of dataset knowledge are foregrounded.

\begin{table}[htbp]
\caption{Percentage distribution of descriptive sub-attribute categories across models and prompt scenarios (T = baseline, TS = title+schema, TSD = title+schema+data)}
\label{tab:distribution}
\begin{center}
\footnotesize
\setlength{\tabcolsep}{4.5pt}   
\renewcommand{\arraystretch}{1.15} 
\begin{tabular}{|p{2.6cm}|c|c|c|c|c|}
\hline
\textbf{Model (Scen.)} & \textbf{Ov.} & \textbf{Cont.} & \textbf{Struct.} & \textbf{Prov.} & \textbf{Clar.} \\
\hline
LLaMA-3 (T) & 21 & 19 & 19 & 15 & 10 \\
LLaMA-3 (TS) & 21 & 22 & 16 & 10 & 10 \\
LLaMA-3 (TSD) & 21 & 22 & 16 & 9  & 13 \\
\hline
Mistral (T) & 33 & 2  & 17 & 0  & 43 \\
Mistral (TS) & 43 & 14 & 6  & 0  & 34 \\
Mistral (TSD) & 39 & 20 & 8  & 0  & 28 \\
\hline
Qwen (T) & 33 & 5  & 16 & 12 & 17 \\
Qwen (TS) & 24 & 16 & 16 & 2  & 27 \\
Qwen (TSD) & 19 & 18 & 12 & 1  & 34 \\
\hline
Gemma (T) & 11 & 13 & 11 & 15 & 24 \\
Gemma (TS) & 10 & 15 & 14 & 14 & 20 \\
Gemma (TSD) & 10 & 14 & 13 & 14 & 20 \\
\hline
\end{tabular}
\end{center}
\end{table}

\subsection{Qualitative Example: Context-Induced Shifts in Description Style}

To complement the quantitative analyses, we present a qualitative example illustrating how progressively richer dataset context reshapes generated descriptions. Fig.~\ref{fig:qual-example} shows excerpted outputs produced by LLaMA-3 for the dataset \emph{``Financial Capability and Child Poverty''} under the baseline and ablation conditions.

\begin{figure}[!htb]
\centering
\fbox{
\begin{minipage}{0.95\columnwidth}
\footnotesize

\textbf{Title-only (T):}  
``This dataset provides insights into the relationship between financial capability and child poverty... It offers a comprehensive understanding of the socio-economic factors influencing the well-being of children from low-income households...''

\vspace{0.6em}
\textbf{Title + Schema (TS):}  
``The dataset consists of two tables... Financial Capability Indicators at the LSOA level... and Geographic Reference Data at the postcode and LSOA levels, including location coordinates and administrative codes...''

\vspace{0.6em}
\textbf{Title + Schema + Data (TSD):}  
``The dataset is comprised of two tables... GFA\_PT0\_RECS to GFA\_PT5\_RECS represent financial participation categories... allowing analysts to identify areas requiring targeted support...''

\end{minipage}
}
\caption{Excerpted LLaMA-3 descriptions for the dataset ``Financial Capability and Child Poverty'' under the baseline and two ablation conditions, illustrating the shift from narrative framing (T), to schema-driven enumeration (TS), to semantically grounded analytical description (TSD).}
\label{fig:qual-example}
\end{figure}

\textbf{Title-only (T).}  
When only the dataset title is provided, the model produces a broadly framed, narrative description emphasizing socio-economic relevance, intended users, and potential applications. The description foregrounds purpose and social context (e.g., supporting policymakers and poverty reduction initiatives), but relies on speculative variable definitions and imagined data sources, reflecting high narrative accessibility but limited grounding.

\textbf{Title + Schema (TS).}  
When schema information is introduced, the description shifts toward a technical, catalogue-oriented style. The model organizes the description around tables, indicators, and geographic reference data, emphasizing structure, linking keys, and variable groupings. While this improves structural specificity, narrative framing and contextual interpretation are reduced, illustrating the schema penalty observed in our LLM-as-a-judge evaluation.

\textbf{Title + Schema + Data (TSD).}  
With the addition of representative data samples, the description becomes more semantically grounded and analytically oriented. The model interprets financial participation categories, distinguishes between LSOA- and postcode-level resolution, and introduces realistic data limitations. This condition partially restores semantic interpretation and content richness, consistent with the data recovery effect, while remaining more structurally focused than the Title-only baseline.

\section{Discussion}

Our results demonstrate that both the amount and type of dataset context fundamentally shape how LLMs construct dataset descriptions. Importantly, richer context does not uniformly improve description quality. Instead, it triggers structural shifts in model behaviour that affect which facets of dataset knowledge are surfaced and which are suppressed.

\subsection{The Schema Paradox: When More Metadata Reduces Description Quality}

The LLM-as-a-judge evaluation reveals a consistent schema penalty across models: providing column headers without data examples often reduces overall description quality. This effect appears to stem from a a tendency to utilise technical terms from the schema or dataset. This displaces narrative framing, provenance cues, and plain-language explanations. The LLMs populate the metadata with technical but unexplained terms (such as 'LSOA').  From a data publishing perspective, this is a critical finding. It suggests that schema alone is not an adequate intermediate signal for generating user-oriented dataset descriptions and may even be counterproductive if not paired with either data samples or explicit narrative prompting.

The partial recovery observed when adding data samples indicates that concrete values help models ground their interpretations, enabling them to recover content richness and structural accuracy. However, even in the full-context condition, provenance and clarity characteristics often remained weaker than in the Title-only baseline. This highlights a design tension: technical inputs improve factual specificity, but risk degrading accessibility unless narrative objectives are explicitly reinforced.

\subsection{Model Personas and Axial Variance in Dataset Knowledge Representation}

The descriptive attribute analysis shows that models exhibit distinct and stable descriptive personas. Gemma acts as a governance adherent, consistently foregrounding administrative lineage and quality cues regardless of context. LLaMA-3 functions as a narrative integrator, incorporating new information while preserving high-level framing. Mistral behaves as a relational specialist, reallocating attention toward structural and entity-centric descriptors. Qwen operates as an analytical scanner, rapidly shifting focus toward granular and insight-oriented statements when exposed to data.

These axial variances imply that model choice is not neutral in automated metadata pipelines. Selecting an LLM determines not only the fluency of generated descriptions, but also the epistemic lens through which dataset knowledge is represented. This has direct implications for portal operators and repository designers: models optimized for governance, human discovery, or technical warehousing will produce systematically different metadata even under identical prompts.

\subsection{Implications for LLM-Supported Data Publishing Workflows}

Taken together, our findings suggest three practical implications. First, data samples are a more reliable grounding signal than schema alone, particularly for improving contents and structure descriptions. Second, provenance and clarity characteristics do not reliably emerge from technical inputs and should be explicitly prompted or separately sourced. Third, automated dataset documentation systems should treat model selection as a core design decision rather than an interchangeable backend choice.
Importantly, our comparison with publisher-provided descriptions reaffirms previous work that LLM-generated metadata already matches or exceeds the quality of real open-data catalogue descriptions in most cases \cite{10.1145/3705328.3748100}. This grounds our ablation results in current practice. We suggest that the core challenge is no longer whether LLMs can generate usable descriptions, but how to shape them to preserve narrative accessibility, provenance cues, and user-centered framing.

From a broader perspective, these results reinforce that dataset description generation is not a purely generative task but a representational one. LLMs do not merely summarise datasets; they interpret what a dataset \emph{is}, and this interpretation varies systematically across models and input conditions.

\subsection{Limitations and Threats to Validity}
Our study has several limitations: we rely on an LLM-as-a-judge framework (which may reflect judge-model biases), use datasets from a single open government portal, and evaluate only a small set of open-weight models under a fixed prompting setup; accordingly, results should be interpreted as comparative trends rather than absolute quality and may not generalize across domains, models, or prompting strategies. Nevertheless, our findings suggest that effective LLM-based documentation depends less on adding more context than on deliberately shaping how dataset knowledge is constructed and presented.

\section{Conclusion}
We examined how dataset context shapes LLM-generated dataset descriptions. Our ablation study shows that more context is not necessarily better: schema alone often degrades narrative quality, and representative data only partially restores technical grounding.

We further find that LLMs exhibit stable, model-specific descriptive profiles, foregrounding different facets of dataset knowledge under identical prompts. This reframes dataset description generation as a representational problem shaped by both context and model choice.

These findings provide concrete guidance for LLM-supported data publishing workflows and motivate future human-centered and cross-domain validation.

\section*{AI-Generated Content Acknowledgement}
We used generative AI tools (ChatGPT, Gemini) to assist with drafting and editing portions of the manuscript and for limited support in generating code. All experimental design, data processing decisions, results, and interpretations were developed, reviewed, and verified by the authors.

\bibliographystyle{IEEEtran}
\bibliography{references}

\end{document}